\documentclass{jpsj-suppl}
\usepackage{txfonts} 

\title{Dibaryons at COSY}

\author{M. \textsc{Bashkanov}$^{1}$, H. \textsc{Clement}$^{2}$, D. P. \textsc{Watts}$^{1}$}

\inst{$^{1}$ School of Physics and Astronomy,
University of Edinburgh, James Clerk Maxwell Building,
Peter Guthrie Tait Road, Edinburgh EH9 3FD, UK  \\
$^{2}$  Physikalisches Institut, Eberhard--Karls--Universit\"at 
 T\"ubingen, Auf der Morgenstelle~14, 72076 T\"ubingen, Germany \\
}

\email{mikhail.bashkanov@ed.ac.uk}

\recdate{August 28, 2015}

\abst{Experiments at the Juelich Cooler Synchrotron (COSY) have now found compelling evidence for a new resonant state in the two-baryon system with mass 2380 MeV and a width of 70 MeV. The structure, containing six valence quarks, constitutes a
so-called dibaryon, either a hexaquark or a hadronic
molecule. The new particle denoted now $d^*(2380)$ has quantum numbers
$I(J^p)=0(3^+)$. The present knowledge about the $d^*$ dibaryon as well
as other implications and possible future experiments are discussed.}

\kword{dibaryon, hexaquark, hadronic molecule}

\begin{document}
\maketitle

\section{Introduction}

Despite their long painful history~\cite{Seth1,Seth2} dibaryon
searches (where "dibaryon" means a state with baryon number $B = 2$ without inference on its internal structure [hexaquark/baryonic-molecule]) have recently received new interest. This has been motivated by the recognition
that there are more complex quark configurations than just the
familiar $q \bar q$ and $qqq$ systems. The "hidden color" aspect makes dibaryons
a particularly interesting object in QCD~\cite{BBC}.

The recent exclusive and kinematically complete measurements performed by the
Wasa-at-COSY collaboration together with the partial wave analysis of the SAID group demonstrated that there is a resonance pole at $(2380\pm 10)-i(40\pm 5))$ MeV in the $^3D_3-^3G_3$ coupled partial waves of $np$ scattering~\cite{MBE1,MBE2,WRK}. This finding matches perfectly with the $I(J^P)=0(3^+)$ resonance structure observed at $\sqrt{s}=2.38$ GeV and width of 70 MeV in the total cross section of four two-pion production reactions ~\cite{mb, MB, MBC, TS1, TS2}. Having revealed the pole in the $np$ scattering amplitudes means that this resonance
structure constitutes a genuine $s-$channel resonance in the system of two baryons. It has been denoted since then by $d^*(2380)$ following the nomenclature used for nucleon excitations. One could naively expect $d^*(2380)$ to be a so-called "deltaron" denoting a deuteron-like bound state of two $\Delta$s. Indeed the decay properties of the $d^*$ dibaryon appear to be driven by the $\Delta-\Delta$ component of the $d^*(2380)$ wave function. However it seems likely that the main part of the $d^*$ wave functions is the hidden color six-quark state, barely coupled to ordinary matter, which could provide the $d^*(2380)$ its narrow width. According to recent evaluations the $d^*(2380)$ spends $2/3$ of its time as a hexaquark and the rest as a $\Delta-\Delta$ molecule~\cite{Huang, Huang1} (note that a perfect $I(J^P)=0(3^+)$ hexaquark would have $4/5$ of the six-quark component and 20 percent of $\Delta-\Delta$~\cite{har}). 

\section{Experimental evidences for a dibaryon resonance}

The golden reaction channel for the observation of the $d^*(2380)$ turned out
to be $pn \to d\pi^0\pi^0$, due to the absence of the isovector background
(present in $pn \to d\pi^+\pi^-$) and only moderate contributions from conventional processes due to $t-$channel Roper and $\Delta\Delta$ excitations~\cite{TS0,TS3,TT}.

Since WASA has been the only detector with a nearly full solid angle coverage
for both charged and neutral particles, which was placed at a hadron accelerator, it is of no surprise that there were no adequate $pn\to d\pi^0\pi^0$ data from previous measurements. Thus it was left to the WASA Collaboration to reveal the pronounced Lorentzian energy dependence sitting upon an only small background in the total cross section of this channel.

Due to the much higher level of conventional two-pion production background in
the non-fusion reactions,  {\it i.e.} $pp\pi^-\pi^0$,  $nn\pi^+\pi^0$,
$pn\pi^0\pi^0$ and $pn\pi^+\pi^-$, a determination of the $d^*$ decay branches
from these reactions is not easy. As well as the interpretation the experimental extraction is challenging, involving neutron in the initial state, final state or both. Nevertheless such measurements could be
successfully performed by the Wasa collaboration. The $d^*$ dibaryon decay
properties were evaluated in Ref. ~\cite{BCS}. They are summarized in Table~1.
\begin{table}
\caption{Experimental branching ratios (BR) of the $d^*(2380)$ resonance
into its decay channels}
\label{tab:1}       
\begin{tabular}{ll}
\hline\noalign{\smallskip}
$d^*$ decay channel & Decay branch \% \\
\noalign{\smallskip}\hline\noalign{\smallskip}

$d\pi^0\pi^0$ & 14(1) \\
$d\pi^+\pi^-$ & 23(2) \\
$pp\pi^-\pi^0$ & 6(1) \\
$nn\pi^+\pi^0$ & 6(1) \\
$pn\pi^0\pi^0$ & 12(2) \\
$pn\pi^+\pi^-$ & 30(5) \\
$pn$ & 12(3) \\
$NN\pi$ & $<10$ \\
\noalign{\smallskip}\hline
\end{tabular}
\end{table}

\section{Structure of the $d^*(2380)$}
All the data ~\cite{MBE1,MBE2, mb, MB, MBC, TS1, TS2} collected so far suggest
that in 88 percent of cases $d^*$ decays into $\Delta\Delta$ and in 12\% to
$pn$ ~\cite{BCS,PBC}. It can be further specified that 90\% of the $pn$ decays
proceed via $^3D_3$ partial wave (angular momentum $L=2$ between nucleons) and
10\% via $^3G_3$ partial wave ($L=4$)~\cite{MBE1,MBE2}. In case of the
$\Delta\Delta$ branch at least 5\% of the decays could be expected to proceed with
two $\Delta$'s in relative $D-wave$ ($L=2$)~\cite{BCS1} - a remarkable feature
for the 80 MeV sub-threshold system. One should also note that the $S-wave$
$\Delta\Delta$ system can not decay into the $L=4$ $pn$ state within one-step,
whereas the $D-wave$ $\Delta\Delta$ part can. So it might be reasonable to
assume that at least 5\% of the $\Delta\Delta$ component in the $d^*$ wave
function is a $D-wave$ $\Delta\Delta$ - very similar to a deuteron with its
5\% of  $D-wave$ $pn$ admixture. The wave function of the $d^*(2380)$ can then
be subdivided into 67\% hexaquark, 31\% $S-wave$ $\Delta\Delta$ and 2\%
$D-wave$ $\Delta\Delta$ configuration.

Very recently Gal and Garcilazo showed that the dynamical process $\Delta\Delta
\to D_{12}\pi \to \Delta\Delta$, where $D_{12}(2150)$ is the $I(J^P)=1(2^+)$
$N\Delta$ state~\cite{AG1,AG2}, leads to an extra attraction in the
$\Delta\Delta$ system and large reduction of the $I(J^P)=0(3^+)$
$\Delta\Delta$ decay width. The amount of $D_{12}\pi$ configuration in the
$d^*(2380)$ is not yet clear. A possible influence of this part on $d^*(2380)$
decay branches still needs to be evaluated. The most promising way to constrain
such a configuration is a $d^* \to NN\pi$ decay measurement. Indeed,  $D_{12}$
has a sizable pionless decay branch $D_{12} \to NN$ nicely seen in $pp \to
D_{12} \to d\pi^+$. So single pion decay of the $d^*$ dibaryon can naturally
arise from the $d^* \to D_{12}\pi \to NN\pi$ process. 

It is absolutely not clear if diquarks play any role in the $d^*$ wave
function. One can imagine $d^*$ as a $\Delta^{++}$ analog, where all
$u-$quarks are substituted by axial-vector $ud$-diquarks. A measurement of the $d^*(2380)$ transition form-factor can help to clarify this possibility.

$d^*(2380)$ formation within Skyrm model was recently studied in Ref. ~\cite{Skyrm}. 

\section{$d^*(2380)$ Photoproduction}
Prior to $d^*$ form-factor measurements one needs to verify another very
important channel - $\gamma d \to d^*$. The reaction $\gamma d \to
d\pi^0\pi^0$ appears to be attractive, since conventional processes are
expected to be particularly small ~\cite{fix,fix1} of the order of only 10 nb
at $T_\gamma=550$ MeV with a smooth energy dependence. The next "best"
two-pion production channel $\gamma d \to d\pi^+\pi^-$ has a background two orders of
magnitude higher with a peak at exactly the position of $d^*$ due
to the Kroll-Ruderman term, unfortunate for $d^*$ photoproduction studies.

A promising way forward is exploit polarisation measurements. The situation in photoproduction looks similar to the one in elastic np scattering: the $d^*(2380)$ resonance contribution
is about 0.17 mb, which is more than two orders below the total elastic cross section. However, with help
of the analysing power, which consists only of interference terms in partial waves, it was possible to filter out reliably the resonance contribution~\cite{MBE1,MBE2}.
The analogous case in photoexcitation of $d^*(2380)$ constitute measurements of the polarisation of the outgoing proton in the reactions $\gamma d \to \vec{p}n$. As in the analysing power of np scattering the angular dependence of the resonance effect in the polarisation of the outgoing proton should be proportional to the associated Legendre polynomial $P^1_3(cos\Theta)$~\cite{MBE2}. Therefore, the maximal resonance effect is expected to be at a scattering angle of $\Theta=90^\circ $. In fact, such an effect has already been looked for previously by Kamae et al. in corresponding data from the Tokyo electron synchrotron ~\cite{TOK1, TOK2}. In order to describe the observed large polarisations in the region of $d^*(2380)$ they fitted a number of resonances to the data, among others also a $J^P = 3^+$ state. However, presumably due to the limited data base they only obtained very large widths for these resonances in the order of 200-300 MeV as one would expect from conventional $\Delta\Delta$ excitations. A new measurement at MAMI will measure final state polarisation of both proton and neutron.

\section{$d^*(2380)$ Electroproduction}

$d^*$ electroproduction appears to be the most promising way to measure the
$d^*$ transition form-factor. One can expect a substantial difference compared
to photoproduction. The simplest way to excite the $d^*$ from a deuteron by a
single photon would be the coupling of the six-quark hidden color component of
the deuteron to the one of $d^*$. Due to the extremely small 6q component of
the deuteron (only $\approx$ 0.15\% ~\cite{Mil}) the $\gamma d \to d^*$
reaction is expected to be highly suppressed. In the case of electroproduction
this reaction would be substituted by  $\gamma^* d \to d^*$, where $\gamma^*$
stands for a virtual photon created in the inelastic electron scattering on
the deuteron. But in addition to single photon exchange one can have
two-photon exchange with  $\gamma^*\gamma^* d \to d^*$. We will lose four
orders of magnitude due to the extra photon but gain again the same four
orders of magnitude by using the $pn$ part of the deuteron wave function
instead of its $6q$ part. In such an unusual case one might have similar strengths of the one- and two-photon excitation cross-sections leading to interesting interference patterns.
Interference effects in $d^*$ electroproduction can enormously improve our knowledge about its internal structure. 
  
\section{$d^*(2380)$ in Nuclei}

Since  the signature of this resonance is observed also in the double-pionic fusion reactions to $^3$He~\cite{MBH,EP} and
$^4$He~\cite{AP}, it obviously is robust enough to survive even in the nuclear enviroment, which may have
interesting consequences for nuclear matter under extreme
conditions, especially for the cases of high density and low temperature. 
The enhancement in the dilepton spectrum observed in heavy-ion
collisions for invariant electron-positron masses in the range 0.15 GeV$ < M_{e^+e^-} <$ 0.6 GeV has recently been traced back to a corresponding
enhancement in pn collisions relative to pp collisions
~\cite{HAD}. Whereas the dilepton spectra from pp collisions are
understood quantitatively, theoretical descriptions fail to account
for the much higher dilepton rate in pn collisions - in particular
regarding the region $ M_{e^+e^-} > 0.3$ GeV  at beam energies
below 2 GeV ("DLS Puzzle" ~\cite{DLS1,DLS2}). In Ref. ~\cite{BC} it
has been shown that the missing strength can be attributed to $\rho^0$
channel $\pi^+\pi^-$ production, see Fig. 5 of Ref.  ~\cite{BC}, which is dominated by conventional
$\Delta\Delta$ excitation due to t-channel meson exchange and
contributions from $d^*(2380)$. The most notable contribution from $d^*(2380)$
on dilepton spectra is expected to happen at $T=1.0-1.2$ GeV/A. The strongest
conventional channel for the $ M_{e^+e-} > 0.3$ GeV dileptons at this energy
is the $\Delta$ resonance expected to be produced rather peripherally leading
to very forward-backward peaking nucleon angular distributions. Both
$d^*(2380)$ and $\Delta\Delta$ dilepton production mechanisms from
Ref.~\cite{BC} are expected to have less anisotropic nucleon angular
distributions. While the $pn \to \Delta\Delta \to e^+e^-pn$ reaction can be
verified at Hades with their $T_n = 1.25$ GeV data, a measurement of the 
$d^*(2380)$ mediated dilepton yield would require somewhat lower energies or
better an energy scan over the $T=1.0-1.2$ GeV/A region.

Dibaryons are bosons, hence not Pauli-blocked, thus allowing for higher densities of compressed nuclear matter. The effect of dibaryons on the equation of state for nuclear matter has been considered in various theoretical investigations, see
e.g. Refs. ~\cite{NM1,NM2,NM3,NM4}. So investigation of the $d^*(2380)$ dibaryon  behaviour in nuclear medium might be an essential step for future neutron star investigations.

\section{Beyond $d^*(2380)$}

There are many predictions of the $d^*(2380)$ related dibaryon companions,
from $d^*_s$ - a heavier member of the $d^*$ SU(3) multiplet to the "mirror
dibaryon" with $I(J^P)=3(0^+)$. While the first one is outside the
WASA-at-COSY detection abilities, the latter one can be
investigated. Considering $d^*(2380)$ as a deltaron with two $\Delta$s spins
aligned, the mirror dibaryon would be a deltaron with anti-aligned spin of
$\Delta$s. Due to its isospin $I=3$ there should be 7 states with charges from
$Z=+4$ to $Z=-2$. Experimentally the most attractive one is the charge $Z=+4$
dibaryon. It has only one decay channel, namely $\Delta^{++}\Delta^{++} \to pp
\pi^+\pi^+$ and can be easily measured. The only problem is that it cannot be
so easily produced. Due to charge and isospin conservation one needs to
produce at least two additional negative pions. In case of proton-proton
collision the reaction $pp \to d^{Z=+4} \pi^-\pi^- \to pp2\pi^+2\pi^-$ is the
minimal option to see such an exotic state. Many theorists have calculated
properties of this resonance candidate starting from the pioneering work of
Dyson and Xuong ~\cite{DYS} back in 1964 until very recently  ~\cite{AG1,AG2},
~\cite{FW,QM2,DBT1,DBT2,DBT3,DBT4}. All calculations predict the mirror
dibaryon to be heavier. However due to absence of the $pn$ decay branch it
could be a bit narrower compared to a $d^*$ of the same mass. The key
observable for this resonance should be a peak in the $M_{pp\pi^+\pi^+}$
spectrum, which would unavoidably create a reflection peak in the
$M_{pp\pi^-\pi^-}$ spectrum. The reflection peak will change its position at
different beam energies, whereas the peak in the $M_{pp\pi^+\pi^+}$ spectrum
should stay in place. So by looking at the $M_{pp\pi^+\pi^+}-M_{pp\pi^-\pi^-}$
difference spectrum at at least two beam energies we might detect the
$d^{Z=+4}$ dibaryon. The main physical background is expected to come from
double-$N^*(1440)$ production. Since the Roper resonance prefers the
$N^*(1440)\to \Delta^{++}\pi^-$ decay over the $N^*(1440)\to \Delta^{0}\pi^+$
decay, such kind of background can also create a false asymmetry in the
$M_{pp\pi^+\pi^+}-M_{pp\pi^-\pi^-}$ difference spectrum, which can be also
disentangled by measuring at two beam energies. The total cross-section for
the $pp \to pp2\pi^+2\pi^-$  in the energy range of interest is expected to be
very small - below $0.7\mu b$ at $T_p = 2.024$ GeV, while the $pp \to pp\pi^+\pi^-\pi^0$ total cross section is $220\mu b$  ~\cite{Pickup}. Three orders of magnitude difference in the total cross-section between  $pp \to pp2\pi^+2\pi^-$ and $pp \to pp\pi^+\pi^-\pi^0$ reactions makes it possible to misidentify $3\pi$ production with $\pi^0$ Dalitz decay as a $4\pi$ production. No sign of a narrow ($\Gamma<150 MeV$) $d^{Z=+4}$ dibaryon was reported so far.
     
\section{Summary and outlook}
After a vast number of unsuccessful searches a non-trivial dibaryon resonance
has now been found and its major decay channels identified. What is missing,
is  a measurement of its electromagnetic form factor, in order to learn about
the size of this object - whether it is of molecular type or a compact
six-quark entity. Further experiments at MAINZ and JLab are expected to resolve this question. Other dibaryon resonance are waiting to be discovered.

\section{Acknowledgement}
We acknowledge valuable discussions with P. Salabura, A. Gal and F. Huang on this issue. This work has been supported by STFC (ST/L00478X/1), BMBF and DFG (CL 214/3-1).

\end{document}